# Role of Coherence in Polarization Response of Hybrid Monolayer MoS$_2$-Gold Nanoparticle Systems

Laura Valencia Molina,*,†,¶,§,△ Zlata Fedorova,*,‡,¶,§,△ Angela Barreda,∥ Maximilian Weissflog,¶,§,⊥ Rocio Camacho Morales,† Seung Heon Han,# Anastasia Romashkina,¶,§ Antony George,#,§ Suprova Das,‡,¶,§,⊥ Ralph Schlegel,@ Andrey Turchanin,#,§ Jeetendra Gour,¶ Thomas Siefke,¶,@ Falk Eilenberger,¶,@ Dragomir Neshev,† and Isabelle Staude‡,¶,§,⊥

†*ARC Centre of Excellence for Transformative Meta-Optical Systems (TMOS), Department of Electronic Materials Engineering, Research School of Physics, The Australian National University, Canberra, ACT 2600, Australia*
‡*Institute of Solid State Physics, Friedrich Schiller University Jena, 07743 Jena, Germany*
¶*Institute of Applied Physics, Friedrich Schiller University Jena, 07745 Jena, Germany*
§*Abbe Center of Photonics, Friedrich Schiller University Jena, 07745 Jena, Germany*
∥*Group of Displays and Photonics Applications, Carlos III University of Madrid, Avda. de la Universidad, 30, Leganés, 28911 Madrid, Spain*
⊥*Max Planck School of Photonics, Hans-Knöll-Straße 1, 07745 Jena, Germany*
#*Institute of Physical Chemistry, Friedrich Schiller University Jena, 07743 Jena, Germany*
@*Fraunhofer Institute for Applied Optics and Precision Engineering IOF, 07745 Jena, Germany*
△ *These authors contributed equally*

E-mail: Laura.ValenciaMolina@anu.edu.au; zlata.fedorova@uni-jena.de

**Abstract**

Monolayers of transition metal dichacogenides show strong second-order nonlinearity and symmetry-driven selection rules from their three-fold lattice symmetry. This process resembles the valley-contrasting selection rules for photoluminescence in these materials. However, the underlying physical mechanisms fundamentally differ since second harmonic generation is a coherent process, whereas photoluminescence is incoherent, leading to distinct interactions with photonic nanoresonators. In this study, we investigate the far-field circular polarization properties of second harmonic generation from $MoS_2$ monolayers resonantly interacting with spherical gold nanoparticles. Our results indicate that the coherence of the second harmonic allows its polarization to be mostly preserved, unlike in an incoherent process, where the polarization is scrambled. These findings provide important insights for future applications in valleytronics and quantum nanooptics, where both coherent and incoherent processes can be probed in such hybrid systems without altering sample geometry or operational wavelength.

Monolayers of transition metal dichalcogenides (1L–TMDs) emerge as promising direct bandgap semiconductors for photonic applications [1,2] exhibiting distinctive electro-optical phenomena such as strong photoluminescence,[3] excitonic effects,[4] tunability by external fields,[5–7] and pronounced nonlinear responses.[8–10] The broken inversion symmetry of a monolayer allows for second harmonic generation (SHG), which is particularly enhanced at excitonic resonances.[11] Owing to the $D_{3h}$ crystal symmetry of 1L–TMDs, the SHG obeys nonlinear circular selection rules, where two left circularly polarized (LCP) photons at the fundamental frequency $\omega$ produce one right circularly polarized (RCP) photon at the double the frequency, and vice versa for opposite handedness.[6,12]

In terms of polarization, this SHG process resembles the valley-locked photoluminescence (PL) in 1L–TMDs, where circularly polarized excitation results in circularly polarized emission.[13,14] Such a PL behavior arises from spin-orbit coupling and broken inversion symmetry, which lock the valleys at the K and K' points of the Brillouin zone to different spin states, giving rise to valley-contrasting optical selection rules.[14,15] This introduces a new binary degree of freedom, opening pathways for applications in classical and quantum information processing.[16,17] Furthermore, the emission wavelength of SHG can be tuned to nearly match that of valley-locked PL, as the exciton ground state (1s) is dipole-allowed for SHG.[18] Despite these similarities, the physical mechanisms underlying SHG and PL are fundamentally different. SHG involves virtual states and no real photon absorption, while PL implies photon absorption and subsequent exciton recombination. This is manifested in distinct coherence properties: SHG is a coherent process,[19] whereas PL is predominantly incoherent.

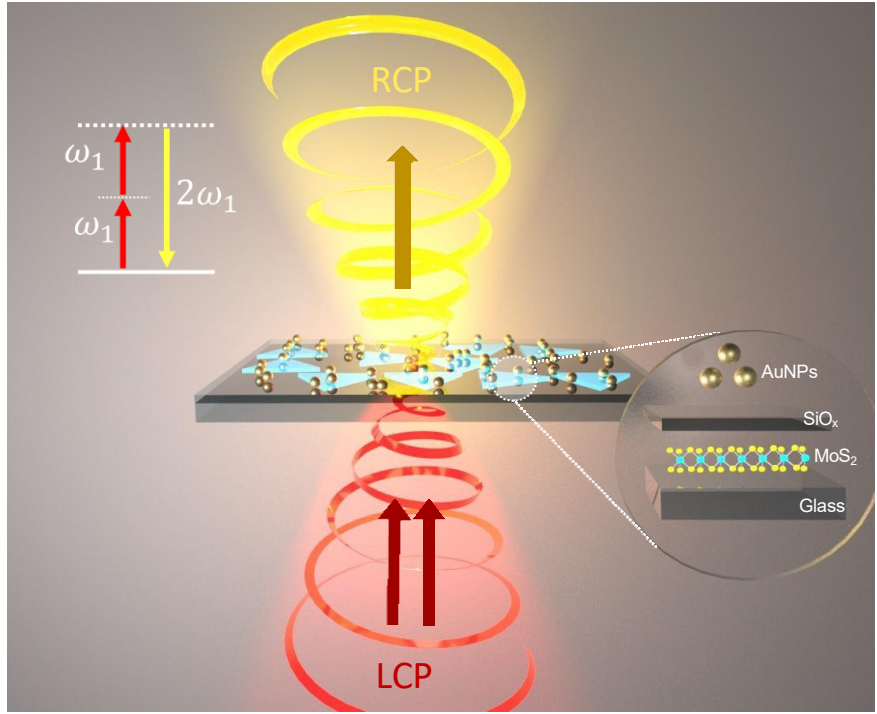

Figure 1: Schematic illustration of the investigated system (not to scale), consisting of the MoS$_2$ monolayers on the glass substrate coated by the thin 5-nm SiO$_x$ layer and decorated by the spherical gold nanoparticles on top. The right inset shows a close-up of the layered structure. The system is illuminated by circularly polarized beam at the fundamental harmonic frequency, while the resulting second harmonic is collected in the transmission configuration. The left inset shows the energy diagram of the SHG process.

Recent efforts to hybridize 1L–TMDs with nanophotonic structures have aimed to enhance control over the valley degree of freedom, focusing on precise addressing, readout, and active manipulation.[20-22] Furthermore, several nanostructures have been shown to effectively modulate the properties of SHG in these materials.[19,23] Despite these advances, fundamen-tal aspects of the light-matter interaction in such hybrid systems remain poorly understood, hindering the development of practical photonic devices.

Namely, while the polarization and wavelength of SHG and valley-locked PL can be tuned to nearly identical conditions, their coherence properties diverge significantly. This contrast, observed under the same geometric configuration, polarization, and wavelength, provides a unique opportunity to explore the role of coherence in such hybrid systems. Understanding how coherence impacts the far-field response of such systems is crucial to advancing nanoscale valleytronic devices.

Previous studies investigated the interaction of valley-selective PL with a resonant gold nanoparticle (AuNP). It was reported that the strong depolarization of circular PL mediated by AuNPs can be attributed to the incoherent nature of the emitters.[24] However, a comprehensive understanding of the behavior in the coherent regime remained unclear. In this work, we address this gap by focusing on coherent circularly polarized SHG at almost the same wavelength. As a simple yet illustrative model system, we consider a $MoS_2$ monolayer resonantly interacting with spherical AuNPs separated by a thin dielectric spacer layer, as shown in Figure 1.

We begin by comparing the behavior of ensembles of coherent rotating dipoles with numerically computed circular SHG emission. This comparison establishes a conceptual link between the linear PL process and the nonlinear SHG response in such hybrid systems (see Figure 2(a)). Building on our numerical findings, we experimentally investigate how AuNPs resonantly interact with a 1L–TMD, influencing the circular polarization properties of the SHG by making use of polarization-resolved SHG microscopy. Our findings reveal a striking contrast between the incoherent and coherent cases. Namely, incoherent circularly-polarized emission is strongly depolarized through the interaction with a nanoparticle, while the degree of circular polarization of the SHG scattered by the nanoparticle remains preserved, mirroring the behavior of coherent dipole ensembles. Notably, in the absence of a nanoparticle, both valley PL and circular SHG exhibit almost perfect circular polarization. This distinct behavior highlights the fundamental role of coherence in shaping the far-field polarization properties of nanoantenna-TMD systems, offering an ideal platform to probe coherence in light-matter interactions with potential applications in classical and quantum nanooptics.

As a stating point we compare two scenarios: incoherent and coherent ensembles of circularly polarized emitters distributed with constant surface density in the *xy*-plane beneath an AuNP at the 1L–TMD's position. The geometry of the simulated area is shown in Figure 2(b). Each RCP (LCP) emitter is modeled as a rotating dipole composed of two perpendicular oscillating point dipoles with a phase offset of $+\frac{\pi}{2}$ ($-\frac{\pi}{2}$), expressed as $\mathbf{p} = (\mathbf{p}_x \pm i\mathbf{p}_y)e^{i\phi}$. In the incoherent case, emitters have random phases $\phi(t)$ uncorrelated in space and time, and the total far-field intensity is obtained by summing the individual intensity contributions over the emitting area, as done in Ref.[24] Conversely, in the coherent case, all emitters radiate with a fixed phase, $\phi$, corresponding to an areal summation of the fields emitted by point sources at different locations with the appropriate space-dependent phase factors. Accordingly, we first compute the emission from individual rotating dipoles at varying radial distances from the AuNP center and then integrate either the intensities or the complex fields, thereby obtaining the total emission for the incoherent and coherent cases, respectively. Using the system's cylindrical symmetry, we simplified the problem by displacing emitters along one axis. To study displacement-dependent behavior, we consider emitters placed on a thin ring of radius $\rho$ centered at the AuNP. The parameters used are motivated by experimental conditions: a 100 nm AuNP radius, a 5 nm dielectric spacer, wavelength of 660 nm, and a refractive index $n$ = 1.5 for both the substrate and spacer. The size-dependent optical properties of the AuNP are derived from the permittivity model in Ref.[25] Details of the calculations are provided in Section S1 of the Supporting Information (SI).

To characterize the polarization state of the emitted light, we calculate the Stokes parameters, with $S_0$ being the total intensity of the SHG. The ratio $S_1/S_0$ corresponds to the linear polarization along the horizontal and vertical axes, $S_2/S_0$ indicates the polarization along the ±45° diagonal axes, and $S_3/S_0$ describes the degree of circular polarization with values ranging from +1 (RCP) to -1 (LCP). Figure 2(c) presents the computed Stokes parameter $S_3/S_0$ for light emitted by coherent and incoherent RCP dipoles as a function of the radial distance $\rho$ from the AuNP's center. Here, dipole-dipole interactions are neglected, such that normalization to $S_0$ removes the dependence on the emitter density. For incoherent dipoles, the $S_3/S_0$ decreases rapidly with distance, reaching a minimum of $-0.13$ at $\rho \approx 30$ nm before returning to unity at larger distances. This behavior arises because the lack of phase correlation among emitters breaks the system's cylindrical symmetry if dipoles are positioned off-center. In contrast, for coherent dipoles, the $S_3/S_0$ remains nearly constant and close to 1, regardless of $\rho$. This is because a ring of in-phase rotating dipoles maintains centrosymmetry, thereby preserving circular polarization. This stark difference underscores the critical role of coherence in shaping the polarization properties of emissions in hybrid systems.

The calculations above demonstrate that circular polarization is preserved only for coherent rotating dipoles. To determine whether the second-harmonic process follows this behavior, we perform numerical simulations of its far-field polarization state upon interaction with an AuNP. The numerical simulations of the SHG response were carried out with the wave optics module in the commercial software COMSOL Multiphysics, which is based on the finite-element method. We use a gold nanoparticle with a 100 nm radius located on top of a silica substrate. An emitting area of size 550 nm by 550 nm was positioned 5 nm below the air-substrate interface, which represents 1L–MoS$_2$. The SHG of the system was calculated by first solving the linear scattering problem with incident fields at the fundamental frequency. Next, based on the fundamental fields, we determined the nonlinear surface current density on the 2D material. Finally, following the undepleted-pump approximation,[26] the linear scattering problem at the second harmonic frequency was solved for the resulting surface current density. Note, that the surface second-harmonic efficiency from an ideal single gold nanoparticle is negligible as a result of its centrosymmetry. Therefore, we only consider $\chi^{(2)}$ for the 1L–MoS$_2$. In the experiment, we use a high-NA objective that collects SHG emission over a large solid angle. Therefore, the experimentally determined Stokes parameters will be an averaged value of the polarization state of potentially many emission directions. To test if the degree of circular polarization has a strong dependence on the collected emission angle, we calculate the polarization resolved momentum distribution ($k$-space image) for our experimental setting following the procedure in Ref.[27] The SHG far field, is hereby obtained using a near-to-far field transformation based on reciprocity arguments (Ref[28]) that handles the combined TMD-nanoresonator system placed on a substrate. The calculated BFP images are shown in Figure 2(d), further details are included in SI Section 2. Note that since $S_1/S_0$ and $S_2/S_0$ have values near zero, there is no significant linear polarization along any of the axes (horizontal/vertical or diagonal/antidiagonal). However, the values of $S_3/S_0$ are close to 1 across all emission directions of light, indicating a uniform RCP state. This finding aligns well with the results obtained for the coherent ensemble of rotating dipoles. For direct comparison with Figure 2(c), we show in Figure S1 (SI) the momentum distributions of the Stokes parameters for coherent rotating dipoles uniformly distributed across a 550 nm diameter circle. The circular polarization remains strongly positive, with values close to unity. Section S2 of SI describes the details of the SHG Stokes parameters for RCP.

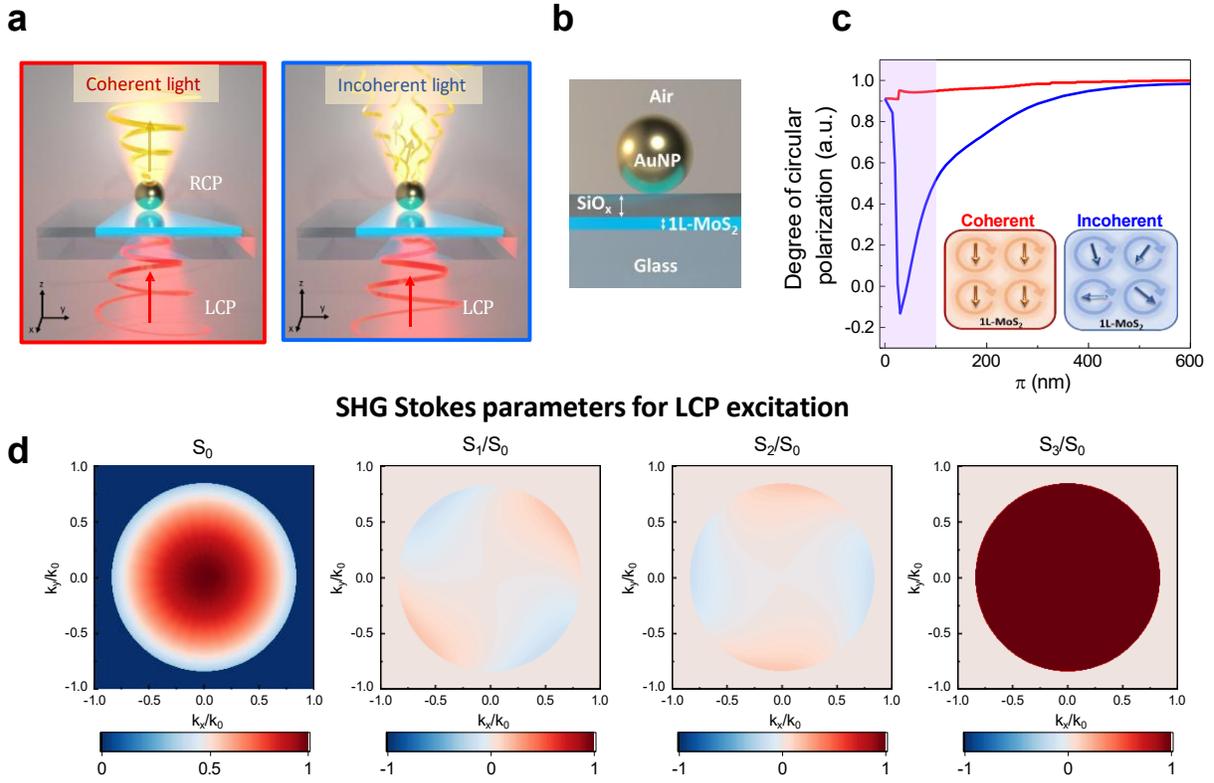

Figure 2: (a) Sketch of the coherent and incoherent process (not to scale). (b) Sketch of the simulated geometry (not to scale). (c) Averaged degree of circular polarization emitted by coherent/incoherent dipoles in $+z$ direction as a function of the distance between the emitter and the AuNP's center. (d) Simulated Stokes parameter SHG momentum space for LCP light, using a NA = 0.85 to match experimental conditions.

For an experimental investigation of the described concepts, we consider a sample consisting of 1L-MoS$_2$ crystals grown by chemical vapor deposition (CVD) method ,[29] then transferred to a glass substrate by wet assisted process. A thin film of 5 nm SiO$_x$ is deposited on top by physical vapor deposition. Finally, AuNPs of 100 nm radius are spin coated on the surface. More details on sample preparation can be found in Section S3 of the SI. To experimentally investigate the polarization response of our hybrid system in the nonlinear regime, we first examine the nonlinear properties of the bare 1L–MoS$_2$ crystals. We use PL spectroscopy to confirm that the CVD-grown MoS$_2$ crystals are monolayers, as indicated by a strong PL signal at direct excitonic transitions, which is absent in multilayer crystals. The PL spectra are corroborated by reflectance measurements showing clear maxima at the spectral positions of the A and B excitons (see Figure S5(a), SI). Furthermore, the scattering spectra for a single AuNP show its resonance overlaping the A- exiton energy of the 1L-MoS$_2$ as shown in Figure S5(b). Importantly, the SHG signal from both the SiO$_x$ substrate and the spherical AuNPs in Figure S7 (SI) remains at the noise level.

We measure the nonlinear emission of the 1L–MoS$_2$ using ∼200-fs pulsed laser at an 80 MHz repetition rate and at constant optical power of 5 mW. The wavelength of the fundmental harmonic (FH) beam is tuned from 1270 nm to 1372 nm displayed in Figure 3(a). At around 1310 nm, when the SHG photon energy ($\bar{h}\omega_{SHG} \simeq$ 1.91 eV) matches the A-excitonic energy of a monolayer MoS$_2$, the SHG response is enhanced, in accord with literature.[30] This is consistent with the spectral position of the A-exciton resonance of the embedded 1L–MoS$_2$ measured by differential reflectance spectroscopy (∼1.89 eV), (see SI, Figure S5). Further-more, we perform power dependence measurements of the SHG signal (see Figure 3(b)) reveiling the expected quadratic dependence on the excitation power. For polarization-resolved analysis, the sample is excited with linearly polarized light and the SHG signal is collected in the same polarization. For normal incidence, a sixfold symmetry in the SHG polarization dependence is expected due to the D$_{3h}$ crystal structure, with SHG intensity varying as I$_{SHG}$ = I$_0$cos$^2$(3($\theta+\theta_0$)), where $\theta$ is the azimuthal angle between the laser polarization and the armchair direction and $\theta_0$ is the offset angle between the armchair direction and the x-axis.[31,32] The SHG polarization dependence measured is shown in Figure 3(c), the dots are the experimental data and the solid line corresponding to the fit function I$_{SHG}$ = I$_0$cos$^2$(3($\theta - \theta_0$)) , where $\theta_0$ = 20° is the offset. Importantly, the observed symmetric sixfold pattern suggests that the sample does not experience any observable strain, preserving the D$_{3h}$ symmetry .[33]

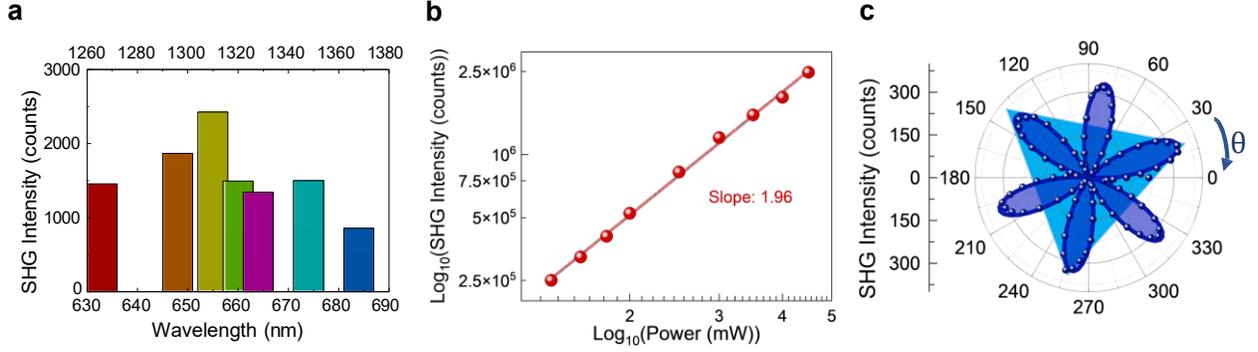

Figure 3: Nonlinear characterization of the CVD-grown MoS$_2$ monolayers. (a) Measured nonlinear response of 1L-MoS$_2$ for a variation of the fundamental harmonic frequency. (b) SHG intensity as a function of average power in the excitation beam (red dots). The experimental data, indicate a quadratic dependence of the SHG processes. (c) Polar plot of the polarization-resolved SHG signal.

Next, we study the effect of the AuNPs on the far-field polarization properties of the circularly polarized SHG from the MoS$_2$ monolayers by polarization-resolved nonlinear microscopy. The polarization of the emitted SHG was measured using the combination of a rotational quarter-wave plate (QWP) and a linear polarizer as shown in Figure 4(a), following the Fourier method.[34] Further details on the measurement scheme are provided in

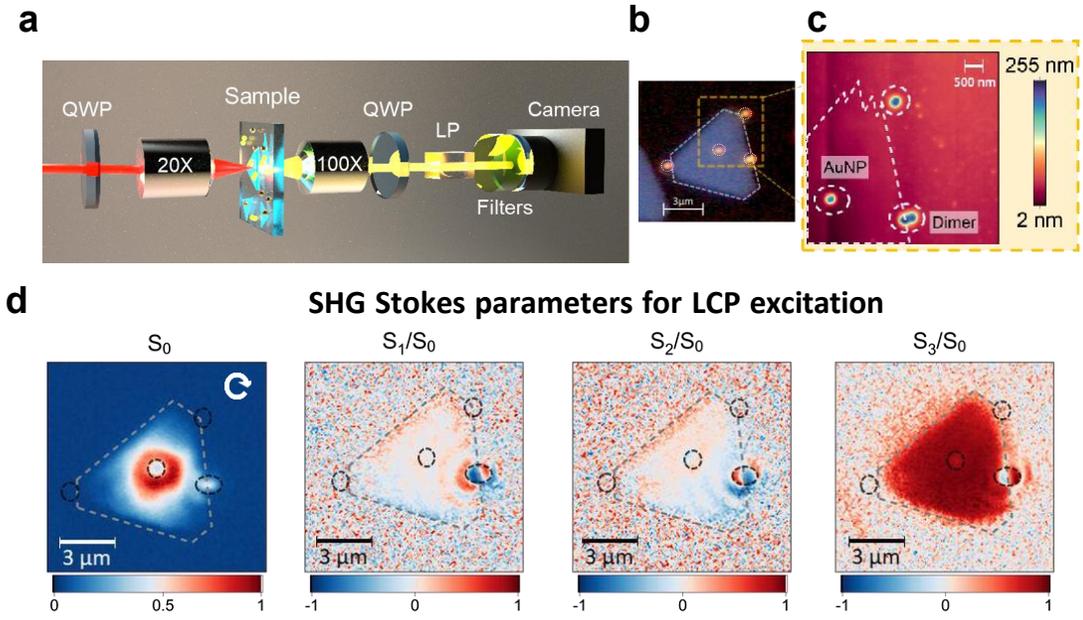

Figure 4: SHG measurements of the hybrid system. (a) Schematic diagram of the SHG experimental setup. Configuration of the measured sample: (b) bright-mode microscope image, and (c) AFM Z-height profile of the sample surface. (d) SHG Stokes parameters: $S_0$ represents SHG total intensity, while the normalized parameters $S_1/S_0$, $S_2/S_0$ and $S_3/S_0$ describe the state of polarization at each point.

Section 5 of the SI. We study the SHG response of 1L-MoS$_2$ crystal with a single AuNP. The configuration of the AuNP on the 1L-MoS$_2$ crystal is shown in Figure 4(b) with the microscope image of the 1L-MoS$_2$ crystal. By examining the Atomic Force Microscope (AFM) images (Figure 4(c)). While several AuNPs can be identified in the region of interest, we first concentrate on the single AuNP located approximately in the centre of the 1L-MoS$_2$ crytsal under investigation. Figure 4(d) shows the normalized Stokes parameters of the SHG signal for an LCP FH beam. Data for the opposite FH handedness (RCP) is provided in Figure S9 of the SI. In both cases, the single AuNP at the center appears as a region of reduced intensity ($S_0$), attributed to resonant absorption at the SHG wavelength. To verify the consistency of this behavior, Figure S8 of the SI shows an SHG scan of a different 1L-MoS$_2$ crystal containing multiple AuNPs, all of which lead to the local signal reduction. We note that measuring the momentum distribution for direct comparison with numerical results was not feasible, as strong absorption at the SHG wavelength makes the nanoparticle emission too weak to distinguish from the unperturbed emission, which dominates the signal. Importantly, Figure 4(d) clearly demonstrates that $S_1/S_0$ and $S_2/S_0$ are close to zero, while $S_3/S_0$ is almost 1, indicating high purity of circular polarization. This is further confirmed by polarization-resolved SHG scans shown in Figure S8. Altogether, our resutls align with the theoretical predictions from nonlinear calculations and the coherent dipole model. The preservation of circular polarization in the presence of a single nanoparticle highlights the fundamental difference between incoherent (linear) and coherent (nonlinear) optical processes.

Finally, to evaluate the sensitivity of our detection method to small polariation changes, we compare the SHG response of a single AuNP with that of a plasmonic dimer at the monolayer's edge (see dashed ellipses in Figure 4 (c)). The tightly confined electromagnetic field in the dimer gap enhances SHG efficiency, an effect extensively studied in both resonant and off-resonant regimes.[35,36] Since the dimer's orientation remains indistinguishable under circularly polarized excitation, Stokes polarimetry offers an additional advantage by detecting and analyzing fine variations in polarization with high precision, thereby enabling the dimer orientation to be clearly visualized. Additional characterization of the linear SHG polarization dependence (SI, Figure S10) further corroborates that the SHG emission lobes are oriented parallel to the dimer axis. Most importantly, these results indicate the ability of our measurement scheme to resolve fine polarization changes confirming its suitability for is particularly valuable in characterizing nanoscale systems with complex interactions.

In this work, we have investigated the polarization response of the SHG from a monolayer $MoS_2$ resonantly interacting with AuNPs. Through polarization-resolved nonlinear microscopy and full-wave simulations, we have demonstrated that the light scattered by the nanoparticle strongly preserves the circular polarization of the SHG emission. This behavior stands in stark contrast to circularly polarized valley-selective photoluminescence, whose polarization is scrambled by the presence of the nanoparticle despite identical sample geometry and emission wavelength.[24]

We have further shown that both processes, SHG and PL, can be described as emissions from an ensemble of rotating dipoles. The key factor leading to their distinct polarization responses is the coherence of the emitters: while SHG maintains a high degree of coherence, PL is largely incoherent. In the absence of the nanoparticle, both processes would lead to the emission of circularly polarized light. However, the nanoparticle's presence selectively preserves the polarization of coherent emission while depolarizing incoherent emission. Intuitively, this follows from symmetry considerations. Without phase correlation, the system's cylindrical symmetry is disrupted when an emitter is positioned off-center relative to the AuNP. In contrast, synchronized emitters preserve this symmetry. This connection between the locally variable polarization response and coherence is crucial for applications where the state of polarization serves as an information carrier.

While our specific system does not enhance the emission, tailored designs, such as plasmonic nanocavities consisting of a nanoparticle coupled to a metallic film, could enable further significant signal enhancement.[37] Such configurations hold promise for controlling and manipulating quantum light sources based on 2D materials. In particular, the development of nanoscale single-photon sources[38] or entangled photon pairs generated via spontaneous parametric down-conversion[39] will greatly benefit from an improved understanding of the interplay between coherence and polarization. Our findings provide valuable insights for the engineering of future quantum photonic devices.

## Acknowledgement

We acknowledge the support by the German Research Foundation DFG (CRC 1375 NOA), project number 398816777 (subprojects C4, B2, B3 and B4); the International Research Training Group (IRTG) 2675 "Meta-Active", project number 437527638; the Federal Ministry for Education and Research (BMBF) project number 16KIS1792 SiNNER and the Australian Research Council via the Centres of Excellence program (CE200100010). . A.B. gratefully acknowledges financial support from the Spanish national project No. PID2022-137857NA-I00. A. B Thanks MICINN for the Ramon y Cajal Fellowship (grant No. RYC2021-030880-I). The authors thank G. Soavi for the useful discussions. The authors are grateful to A. Fedotova and F. Löchner for their help with setting up the nonlinear experiment. The authors would like to thank the microstructure technology team at IAP Jena for the fabrication of the Cr grid patterns.

## Supporting Information Available

Numerical details of the dipole rotation emission simulation, SHG simulation, SHG Stokes parameter calculation, as well as additional information on sample preparation, differential reflectance, 1L-$MoS_2$ PL, simulated and measured scattering for a single AuNP, and nonlinear characterization setup including SHG Stokes parameters and SHG polarimetry.

## References

(1) Mak, K. F.; Shan, J. Photonics and optoelectronics of 2D semiconductor transition metal dichalcogenides. *Nature Photonics* **2016**, *10*, 216–226.

(2) Datta, I.; Chae, S. H.; Bhatt, G. R.; Tadayon, M. A.; Li, B.; Yu, Y.; Park, C.; Park, J.; Cao, L.; Basov, D.; others Low-loss composite photonic platform based on 2D semiconductor monolayers. *Nature Photonics* **2020**, *14*, 256–262.

(3) Mak, K. F.; Lee, C.; Hone, J.; Shan, J.; Heinz, T. F. Atomically thin MoS 2: a new direct-gap semiconductor. *Physical review letters* **2010**, *105*, 136805.

(4) Chernikov, A.; Berkelbach, T. C.; Hill, H. M.; Rigosi, A.; Li, Y.; Aslan, B.; Reichman, D. R.; Hybertsen, M. S.; Heinz, T. F. Exciton binding energy and nonhydrogenic Rydberg series in monolayer WS 2. *Physical review letters* **2014**, *113*, 076802.

(5) Ross, J. S.; Wu, S.; Yu, H.; Ghimire, N. J.; Jones, A. M.; Aivazian, G.; Yan, J.; Mandrus, D. G.; Xiao, D.; Yao, W.; others Electrical control of neutral and charged excitons in a monolayer semiconductor. *Nature communications* **2013**, *4*, 1474.

(6) Seyler, K. L.; Schaibley, J. R.; Gong, P.; Rivera, P.; Jones, A. M.; Wu, S.; Yan, J.; Mandrus, D. G.; Yao, W.; Xu, X. Electrical control of second-harmonic generation in a WSe2 monolayer transistor. *Nature nanotechnology* **2015**, *10*, 407–411.

(7) MacNeill, D.; Heikes, C.; Mak, K. F.; Anderson, Z.; Kormányos, A.; Zólyomi, V.; Park, J.; Ralph, D. C. Breaking of valley degeneracy by magnetic field in monolayer MoSe 2. *Physical review letters* **2015**, *114*, 037401.

(8) Autere, A.; Jussila, H.; Dai, Y.; Wang, Y.; Lipsanen, H.; Sun, Z. Nonlinear optics with 2D layered materials. *Advanced Materials* **2018**, *30*, 1705963.

(9) Zhou, R.; Krasnok, A.; Hussain, N.; Yang, S.; Ullah, K. Controlling the harmonic generation in transition metal dichalcogenides and their heterostructures. *Nanophotonics* **2022**, *11*, 3007–3034.

(10) Klimmer, S.; Lettau, T.; Molina, L. V.; Kartashov, D.; Peschel, U.; Wilhelm, J.; Neshev, D.; Soavi, G. Ultrafast Coherent Bandgap Modulation Probed by Parametric Nonlinear Optics. 2025; https://arxiv.org/abs/2504.06130.

(11) Wang, G.; Marie, X.; Gerber, I.; Amand, T.; Lagarde, D.; Bouet, L.; Vidal, M.; Balocchi, A.; Urbaszek, B. Giant enhancement of the optical second-harmonic emission of WSe 2 monolayers by laser excitation at exciton resonances. *Physical review letters* **2015**, *114*, 097403.

(12) Simon, H. J.; Bloembergen, N. Second-Harmonic Light Generation in Crystals with Natural Optical Activity. *Phys. Rev.* **1968**, *171*, 1104–1114.

(13) Xiao, D.; Liu, G.-B.; Feng, W.; Xu, X.; Yao, W. Coupled spin and valley physics in monolayers of MoS 2 and other group-VI dichalcogenides. *Physical review letters* **2012**, *108*, 196802.

(14) Mak, K. F.; He, K.; Shan, J.; Heinz, T. F. Control of valley polarization in monolayer $MoS_2$ by optical helicity. *Nature Nanotechnology* **2012**, *7*, 494–498.

(15) Zeng, H.; Dai, J.; Yao, W.; Xiao, D.; Cui, X. Valley polarization in $MoS_2$ monolayers by optical pumping. *Nature Nanotechnology* **2012**, *7*, 490–493.

(16) Schaibley, J. R.; Yu, H.; Clark, G.; Rivera, P.; Ross, J. S.; Seyler, K. L.; Yao, W.; Xu, X. Valleytronics in 2D materials. *Nature Reviews Materials* **2016**, *1*, 1–15.

(17) Balla, N. K.; O'Brien, M.; McEvoy, N.; Duesberg, G. S.; Rigneault, H.; Brasselet, S.; McCloskey, D. Effects of Excitonic Resonance on Second and Third Order Nonlinear Scattering from Few-Layer MoS2. *ACS Photonics* **2018**, *5*, 1235–1240.

(18) Herrmann, P.; Klimmer, S.; Lettau, T.; Monfared, M.; Staude, I.; Paradisanos, I.; Peschel, U.; Soavi, G. Nonlinear All-Optical Coherent Generation and Read-Out of Valleys in Atomically Thin Semiconductors. *Small* **2023**, *19*, 2301126.

(19) Hu, G.; Hong, X.; Wang, K.; Wu, J.; Xu, H.-X.; Zhao, W.; Liu, W.; Zhang, S.; Garcia-Vidal, F.; Wang, B.; others Coherent steering of nonlinear chiral valley photons with a synthetic Au–WS2 metasurface. *Nature Photonics* **2019**, *13*, 467–472.

(20) Wang, J.; Li, H.; Ma, Y.; Zhao, M.; Liu, W.; Wang, B.; Wu, S.; Liu, X.; Shi, L.; Jiang, T.; others Routing valley exciton emission of a WS2 monolayer via delocalized Bloch modes of in-plane inversion-symmetry-broken photonic crystal slabs. *Light: Science & Applications* **2020**, *9*, 148.

(21) Zheng, L.; Dang, Z.; Ding, D.; Liu, Z.; Dai, Y.; Lu, J.; Fang, Z. Electron-Induced

Chirality-Selective Routing of Valley Photons via Metallic Nanostructure. *Advanced Materials* **2023**, *35*, 2204908.

(22) Liu, Y.; Lau, S. C.; Cheng, W.-H.; Johnson, A.; Li, Q.; Simmerman, E.; Karni, O.; Hu, J.; Liu, F.; Brongersma, M. L.; others Controlling Valley-Specific Light Emission from Monolayer MoS2 with Achiral Dielectric Metasurfaces. *Nano letters* **2023**, *23*, 6124–6131.

(23) Li, C.; Lu, X.; Srivastava, A.; Storm, S. D.; Gelfand, R.; Pelton, M.; Sukharev, M.; Harutyunyan, H. Second harmonic generation from a single plasmonic nanorod strongly coupled to a WSe2 monolayer. *Nano Letters* **2020**, *21*, 1599–1605.

(24) Bucher, T.; Fedorova, Z.; Abasifard, M.; Mupparapu, R.; Wurdack, M. J.; Najafidehaghani, E.; Gan, Z.; Knopf, H.; George, A.; Eilenberger, F.; others Influence of resonant plasmonic nanoparticles on optically accessing the valley degree of freedom in 2D semiconductors. *Nature Communications* **2024**, *15*, 10098.

(25) Derkachova, A.; Kolwas, K.; Demchenko, I. Dielectric function for gold in plasmonics applications: size dependence of plasmon resonance frequencies and damping rates for nanospheres. *Plasmonics* **2016**, *11*, 941–951.

(26) Celebrano, M.; Wu, X.; Baselli, M.; Großmann, S.; Biagioni, P.; Locatelli, A.; Angelis, C. D.; Cerullo, G.; Osellame, R.; Hecht, B.; Duò, L.; Ciccacci, F.; Finazzi, M. Mode matching in multiresonant plasmonic nanoantennas for enhanced second harmonic generation. *Nature Nanotechnology* **2015**, *10*, 412–417.

(27) Weissflog, M. A. et al. Far-Field Polarization Engineering from Nonlinear Nanoresonators. *Laser & Photonics Reviews* **2022**, *16*, 2200183.

(28) Yang, J.; Hugonin, J.-P.; Lalanne, P. Near-to-Far Field Transformations for Radiative and Guided Waves. *ACS Photonics* **2016**, *3*, 395–402, doi: 10.1021/acsphotonics.5b00559.

(29) George, A.; Neumann, C.; Kaiser, D.; Mupparapu, R.; Lehnert, T.; Hübner, U.; Tang, Z.; Winter, A.; Kaiser, U.; Staude, I. Controlled growth of transition metal dichalcogenide monolayers using Knudsen-type effusion cells for the precursors. *Journal of Physics: Materials* **2019**, *2*, 016001.

(30) Klimmer, S.; Ghaebi, O.; Gan, Z.; George, A.; Turchanin, A.; Cerullo, G.; Soavi, G. All-optical polarization and amplitude modulation of second-harmonic generation in atomically thin semiconductors. *Nature Photonics* **2021**, *15*, 837–842.

(31) Malard, L. M.; Alencar, T. V.; Barboza, A. P. M.; Mak, K. F.; de Paula, A. M. Observation of intense second harmonic generation from $MoS_2$ atomic crystals. *Phys. Rev. B* **2013**, *87*, 201401.

(32) Mennel, L.; Paur, M.; Mueller, T. Second harmonic generation in strained transition metal dichalcogenide monolayers: MoS2, MoSe2, WS2, and WSe2. *APL Photonics* **2018**, *4*, 034404.

(33) Psilodimitrakopoulos, S.; Ilin, S.; Zelenkov, L. E.; Makarov, S.; Stratakis, E. Tailoring of the polarization-resolved second harmonic generation in two-dimensional semiconductors. *Nanophotonics* **2024**, *13*, 3181–3206.

(34) Schaefer, B.; Collett, E.; Smyth, R.; Barrett, D.; Fraher, B. Measuring the Stokes polarization parameters. *American Journal of Physics* **2007**, *75*, 163–168.

(35) Slablab, A.; Le Xuan, L.; Zielinski, M.; De Wilde, Y.; Jacques, V.; Chauvat, D.; Roch, J.-F. Second-harmonic generation from coupled plasmon modes in a single dimer of gold nanospheres. *Optics express* **2011**, *20*, 220–227.

(36) Wang, Y.; Peng, Z.; De Wilde, Y.; Lei, D. Symmetry-breaking-induced off-resonance second-harmonic generation enhancement in asymmetric plasmonic nanoparticle dimers. *Nanophotonics* **2024**, *13*, 3337–3346.

(37) Han, X.; Wang, K.; Persaud, P. D.; Xing, X.; Liu, W.; Long, H.; Li, F.; Wang, B.; Singh, M. R.; Lu, P. Harmonic resonance enhanced second-harmonic generation in the monolayer WS2–Ag nanocavity. *Acs Photonics* **2020**, *7*, 562–568.

(38) Lee, S.-J.; So, J.-P.; Kim, R. M.; Kim, K.-H.; Rha, H.-H.; Na, G.; Han, J. H.; Jeong, K.-Y.; Nam, K. T.; Park, H.-G. Spin angular momentum–encoded single-photon emitters in a chiral nanoparticle–coupled WSe2 monolayer. *Science Advances* **2024**, *10*, eadn7210.

(39) Weissflog, M. A. et al. A Tunable Transition Metal Dichalcogenide Entangled Photon-Pair Source. *Nature Communications* **2024**, *15*, 7600.